\documentclass[]{article}
\usepackage{graphicx}
\usepackage{subfigure}
\usepackage{natbib}
\usepackage{amssymb,amsmath}
\bibpunct{(}{)}{;}{a}{,}{,}
\newcommand{\E}{_{\mbox{\scriptsize E}}}

\newcommand{\sstar}{_{\star}}

\newcommand{\CME}{_{\mbox{\scriptsize CME}}}

\begin{document}

\title{Stellar activity and magnetic shielding}

\author{
J.-M. Grie{\ss}meier\\
ASTRON, Postbus 2, 7990 AA, Dwingeloo, The Netherlands\\
\and
M. Khodachenko\\
Space Research Institute, Austrian Academy of Sciences, \\
Schmiedlstr. 6, A-8042 Graz, Austria\\
\and
H. Lammer\\
Space Research Institute, Austrian Academy of Sciences, \\
Schmiedlstr. 6, A-8042 Graz, Austria\\
\and 
J. L. Grenfell\\
Zentrum f\"ur Astronomie und Astrophysik,
\\Technische Universit\"{a}t Berlin (TUB), \\ 
Hardenbergstr. 36, 10623 Berlin, Germany\\
\and 
A. Stadelmann\\
Technical University of Braunschweig, \\
Mendelssohnstra{\ss}e 3, 38106 Braunschweig, Germany
\and 
U. Motschmann\\
Technical University of Braunschweig, \\
Mendelssohnstra{\ss}e 3, 38106 Braunschweig, Germany
}



\date{}
\maketitle

\begin{abstract}
Stellar activity has a particularly strong influence on planets at small orbital distances, such
as close-in exoplanets. For such planets, we present two extreme cases of stellar
variability, namely stellar coronal mass ejections and stellar wind, 
which both result in the planetary environment being variable on a timescale
of billions of years. For both cases, direct interaction of the streaming plasma with the
planetary atmosphere would entail servere consequences.
In certain cases, however, the planetary atmosphere can be effectively shielded by a strong
planetary magnetic field. The efficiency of this shielding is determined by the planetary
magnetic dipole moment, which is difficult to constrain by either models or observations.
We present different factors which influence the strength of the planetary magnetic dipole
moment. 
Implications are discussed, including nonthermal atmospheric loss,
atmospheric biomarkers, and planetary habitability.
\end{abstract}

\section{Introduction}

One of the many fascinating questions in the field of exoplanet studies is the search for 
habitable worlds.
Because of their relatively small mass, low luminosity, long lifetime and large abundance in the
galaxy, M dwarfs are sometimes suggested as prime targets in searches for terrestrial habitable 
planets \citep{Tarter07, Scalo07}.
Interestingly, M dwarfs also seem to have a larger number of Super-Earth planets 
(i.e.~planets with a mass smaller than 10 terrestrial masses, i.e.~$M \lesssim 10 M\E$)
than more massive stars \citep{Forveille09}.
Excluding planets orbiting around pulsars, 
the first super-Earth detected was GJ 876d, a planet with $\sim\!\!7.5 M\E$ 
or\nolinebreak[4]biting an M star \citep{Rivera05}. 
Thereafter, other super-Earth planets have been discovered 
\citep{Beaulieu06,Lovis06,Udry07,Ribas08,Mayor09,Forveille09,Howard09,Bouchy09}.
In total, 12 planets with masses $\lesssim 10 M\E$ are reported today, and more detections are
expected for the near future \citep{Howard09,Mayor09}\footnote{Up-to-date numbers can be found at the Extrasolar Planets 
Encyclopaedia: {\tt http://www.exoplanet.eu}}. 
The least massive planet has a mass of 4.2 $M\E$.

For M stars the habitable zone \citep[HZ, defined by the range of orbital distances
over which liquid water is possible on the planetary surface, see e.g.][]{Kasting93}
is much closer to the star (depending on stellar mass, but typically $\le$0.3 AU).  
Such close-in distances pose unique problems and constraints to habitability. 
For example, such planets are exposed to strong stellar winds in the early phases of the host star's evolution.
Also,
one expects a large number of 
coronal mass ejections (CMEs) on active M stars \citep{Houdebine90,Scalo07}. 
The related interaction of the dense plasma flux (either from the stellar wind or a CME) with the 
atmosphere/magnetosphere environment of the exposed planets during the active 
stage of the stellar evolution could be strong enough to erode the atmosphere 
or the planets' water inventory via
non-thermal atmospheric loss processes \citep{Khodachenko07AB, Lammer07AB}. 
Weak magnetic shielding of tidally locked planets will also lead to an increased influx of 
galactic cosmic rays and stellar energetic particles \citep{Griessmeier05a,Griessmeier09}, which can have 
important consequences for biological systems, both directly and indirectly via the modification 
of the planetary atmosphere  
\citep[e.g.~destruction of atmospheric ozone, see][]{Grenfell07AB,Grenfell09}.

This paper is organised as follows:
The evolution of the conditions in the stellar vicinity over the
timescale of the stellar evolution is presented in Section
\ref{sec:stellarvar}. Two cases are considered: stellar CMEs (Section \ref{sec:stellarvar:CME}), 
and the average stellar wind (Section \ref{sec:stellarvar:sw}).
Planetary magnetic shielding against the influence of the star is 
discussed in Section \ref{sec:magneticshield}. 
Consquences are discussed in Section \ref{sec:consequences}, with special focus on the potential evaporation of the
planetary atmosphere (Section \ref{sec:consequences:atm}) and the flux of cosmic rays to the planetary atmosphere
(Section \ref{sec:consequences:cr}).
Section \ref{sec:conclusions} closes with a few concluding remarks.

\section{Stellar variability}\label{sec:stellarvar}

Stars are known to be variable in may different ways and on many different timescales.
Effects such as stellar CMEs, flares, starspots and stellar rotation occur on timescales of 
hours or days, while stellar magnetic cycles are expected on the timescale of decades or even 
centuries. Stellar irradiation, the stellar CME activity and the stellar wind parameters evolve 
very slowly over Gyr.
Many of these effects are discussed elsewhere in this volume. 
In this section we focus on the slow evolution of two of the stellar parameters over the stellar 
evolutionary timescale, namely the evolution of stellar CME parameters and of the average 
stellar wind.

\subsection{Stellar CMEs}\label{sec:stellarvar:CME}

There are two reasons why in addition to the steady stellar wind CMEs have to be taken into 
account for the interaction with the planetary atmosphere: (a) Due to their limited propagation
distance, CMEs are more frequent for close-in exoplanets than at larger distances, and 
(b) CMEs can be much denser 
and faster than the average stellar wind, thus exerting a considerable influence on the planet.

At 1 AU, CME collisions with the terrestrial magnetosphere are relatively rare events.
However, this is due to the fact that most CMEs cannot be detected at orbital distances $\ge
0.1$ AU. Only $\le$20\% of CMEs can reach distances of $\approx 1$ AU and more.
Thus, for a planet at 0.1 AU, where all CMEs have to be taken into account, the number of CME-planet 
collisions is expected to be considerably higher \citep{Khodachenko07AB}.

An estimation of CME parameters 
\citep[combining in-situ measurements near the sun (e.g.~by Helios) with remote solar 
observation by SoHO,][]{Khodachenko07AB} has shown that for small orbital
distances, CMEs can be considerably denser and faster than the solar wind (roughly one order of 
magnitude denser and half an order of magnitude faster). Here, we assume that stellar CMEs behave 
similarly to solar ones \citep[``solar-stellar analogy'', see][]{Khodachenko07AB}. 
If CME-planet collisions are frequent enough (such that the planet effectively moves through a 
constant background of CME plasma), the 
interaction of the planetary atmosphere with stellar CMEs can be estimated by replacing
the stellar wind parameters $n(d)$ and $v(d)$ by the corresponding parameters $n\CME(d)$ and 
$v\CME(d)$ for strong CMEs as given by \citet{Khodachenko07AB}.
The consequences for atmospheric erosion will be discussed in Section \ref{sec:consequences:atm}.

\subsection{Stellar wind}\label{sec:stellarvar:sw}

The stellar wind density $n$ and velocity $v$ encountered by a planet are key 
parameters defining the size of the magnetosphere.
As these stellar wind parameters depend on the stellar age, 
the quality of magnetic shielding is strongly time dependent.

For stellar ages $>0.7$ Gyr, the
radial dependence of the stellar wind properties is well described by a
Parker-like \citep{Parker58} stellar wind model. 
In this model, the interplay between stellar gravitation and pressure gradients leads to 
a supersonic gas flow for sufficiently large substellar distances $d$. 
The free parameters are the coronal temperature
and the stellar mass loss. They are indirectly chosen by setting the stellar wind conditions at 1 AU.
More details on the model 
can be found elsewhere \citep[e.g.][]{Parker58,Mann99,Proelss04,GriessmeierPSS06}. 

\begin{figure}
\begin{center}
 	\subfigure[Stellar wind density $n(1 \mbox{AU}, t)$.]{
 	\includegraphics[width=0.48\linewidth]{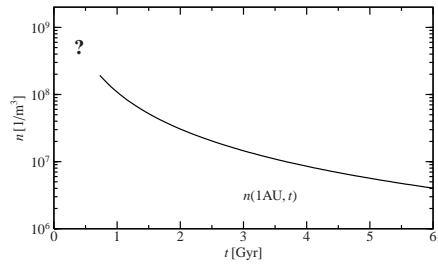} 
   	\label{fig:swTn}}
   	\vspace{1.0cm}
 	\subfigure[Stellar wind velocity $v(1 \mbox{AU}, t)$.]{
 	\includegraphics[width=0.48\linewidth]{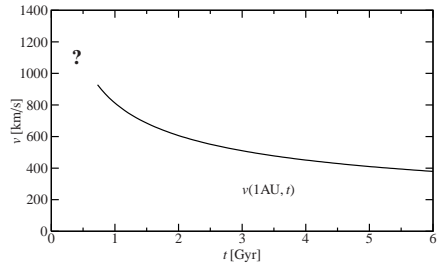} 
	\label{fig:swTv}}
\caption{\label{fig:swT}
Stellar wind density $n(t)$ and velocity $v(t)$ at 1 AU distance of a sun-like star 
 	($M_\star=M_\odot$, $R_\star=R_\odot$), as a function of stellar age.}
\end{center}
\end{figure}

The dependence of the stellar wind density $n$ and velocity and $v$ on the age of the
stellar system is measured by observing astrospheric absorption features of stars with
different ages.
The characteristic Ly$\alpha$ absorption feature (at 1216 $\mbox{\AA}$) created by 
neutral hydrogen at the astropause
was detected in high-resolution 
observations obtained by the Hubble Space Telescope (HST). 
Comparing the measured absorption to that calculated by hydrodynamic codes, these measurements 
allowed the first empirical estimation of the evolution of the stellar mass loss rate as a 
function of stellar age \citep{Wood02,Wood04,Wood05}. 
It should be noted, however, that the resulting estimates are only valid for stellar ages $t\ge0.7$ 
Gyr \citep{Wood05}.
These results for the mass loss rate can be combined with the model for the  
age-dependence of the stellar wind velocity of  \citet{Newkirk80} 
to obtain the age-dependent stellar wind density.
One thus finds \citep{GriessmeierPSS06}:
\begin{equation}
  v(1 \mbox{AU}, t)= v_0 \left( 1+\frac{t}{\tau}\right)^{-0.43},
  \label{eq:scaleV}
\end{equation}
and
\begin{equation}
  n(1 \mbox{AU}, t)= n_0 \left( 1+\frac{t}{\tau}\right)^{-1.86\pm0.6},
  \label{eq:scaleN}
\end{equation}
with 
$v_0=3971$ km/s, $n_0=1.04\cdot10^{11}\mbox{ m}^{-3}$ and
$\tau=2.56\cdot10^7 \mbox{yr}$. 
The dependence of $n(1 \mbox{AU}, t)$ and $v(1 \mbox{AU}, t)$ as a function of stellar age is
shown in Figs.~\ref{fig:swTn} and \ref{fig:swTv} for the case of a sun-like star.

The procedure to obtain the stellar wind velocity 
$v(d,M_\star,R_\star)$ and density $n(d,M_\star,R_\star)$ at the location of an 
exoplanet (i.e.~at distance
$d$) for a host star of given mass $M_\star$, and radius $R_\star$ and age $t$
from $v(1 \mbox{AU}, t)$ and $n(1 \mbox{AU}, t)$
is described in \citet{Griessmeier05a,Griessmeier09}.

\section{Magnetic shielding}\label{sec:magneticshield}

For an efficient magnetic shield to exist, the planet requires a strong magnetic field.
A necessary condition for a magnetic dynamo is a large, electrically conducting fluid region in 
non-uniform motion (i.e. the liquid outer core for terrestrial planets, or a layer of electrically
conducting hydrogen for gas giants). 
According to planetary dynamo theory, this flow should be convective in nature \citep{Stevenson83}. 
In practice, all planets with core convection can be assumed to have a 
magnetic dynamo \citep{Stevenson83}. Conversely, convection can be regarded as a necessary
requirement for a planetary magnetic field. 

Convection is assumed to require plate tectonics in order to operate. The reason for this is that 
plate tectonics helps to cool the planetary interior more efficiently than a stagnant lid
configuration \citep{Stevenson03}.
Efficient cooling is required to maintain convection, and is thus essential to keep the action of a
magnetic dynamo \citep{Stevenson83}. The lack of plate tectonics is also suggested as one of the
possible reasons for the lack of an intrinsic magnetic field on Venus and Mars
\citep{Stevenson83b,Stevenson03}. Thus, plate tectonics may be required to keep an Earth-like
magnetic dynamo in operation over geological time spans \citep{Lammer09AARV}.

There are two reasons why the requirement for plate tectonics for convection might be important as
far as exoplanets are concerned: (a) For close-in exoplanets, the small orbital distance usually
means that the planet is tidally locked \citep[e.g.][]{Griessmeier09}. It is currently not clear how
this influences plate tectonics, but it is sometimes suggested that tidal locking may inhibit plate
tectonics partially or completely. (b) For Super-Earth planets, is it still being debated whether
plate tectonics is likely to occur or not \citep{Valencia07c,ONeill07}. 

Thus, we see that plate tectonics is essential for convection, which is a necessary ingredient for 
magnetic dynamo action. In the following, we will assume that a magnetic dynamo does exist.
However, even if convection is present and a magnetic dynamo exists, the magnetic moment 
is not necessarily large enough to efficiently shield the planet.
The tidally locked state of close-in exoplanets means that they have low rotation rates,
which -- according to scaling laws derived from theoretical
considerations -- leads to a reduction in 
the planetary magnetic moment \citep[e.g.][]{Griessmeier05a}
\footnote{While the analytical models considered here predict a decrease of
magnetic moment with decreasing rotation rate, numerical
experiments indicate that the magnetic moment may be independent of the
angular frequency. 
\citet{Christensen06} and \citet{Olson06} have studied numerically
an extensive set of dynamo models, 
varying the relevant control parameters by at least two orders of magnitude.
In these simulations, the magnetic field is mainly
controlled by the buoyancy flux.
This is in contradiction with the previously discussed concept.
Future studies are required to clarify the relation between
magnetic field strength, the planetary rotation rate,
electrical conductivity, and buoyancy flux.}.

Following the approach of \citet{Griessmeier05a}, one finds that 
for the slowly rotating tidally locked planets, 
the magnetic moment is much 
smaller than for freely rotating planets. For example, an Earth-like planet in an orbit 
of 0.2 AU around a star with 0.5 stellar masses, may have a magnetic moment in the range 
$0.02 \mathcal{M}\E<\mathcal{M}<0.15 \mathcal{M}\E$ ($\mathcal{M}\E$ is 
the value of Earth's current magnetic moment). 
In the following, the minimum value will be adopted to 
estimate an upper limit for atmospheric espace (Sect.~\ref{sec:consequences:atm}), 
but for the cosmic ray flux (Sect.~\ref{sec:consequences:cr}), $\mathcal{M}$ is assued
to take the maximum allowed value.

In the following, the magnetosphere is modelled as a cylinder topped by a half-sphere 
\citep{Voigt81,Stadelmann04,Griessmeier05a}. 
The size of the
magnetosphere is characterized by the magnetopause standoff distance $R_s$, i.e. the
extent of the magnetosphere along the line connecting the star and the planet.
$R_s$ can be obtained from the pressure equilibrium at the
substellar point. This pressure balance includes the stellar wind ram pressure, the stellar wind
thermal pressure of electrons and protons, and the planetary magnetic field pressure.
For a given planetary orbital distance $d$, only the magnetospheric magnetic pressure is a 
function of the distance to the planet, while the other contributions are constant. 
Thus, 
the standoff distance $R_s$ is found to be \citep{GriessmeierPSS06}:
\begin{equation}
	 R_s = 
	 \left[ \frac{\mu_0f_0^2\mathcal{M}^2}
	 {8\pi^2 \left(m n v^2+2 \, nk_BT\right)} \right]^{1/6}.
	 \label{eq:Rs}
\end{equation}
Here, $f_0=1.16$ is the form factor of the magnetosphere and includes the magnetic field caused
by the currents flowing on the magnetopause \citep{Voigt95,Griessmeier04}. From eq.~(\ref{eq:Rs}) it
is evident that a small magnetic moment $\mathcal{M}$, a high stellar wind velocity $v$ or a high
stellar wind density $n$ will lead to a small magnetosphere.

\section{Consequences}\label{sec:consequences}

The weak magnetic shielding of close-in planets against stellar wind of young stars and against
stellar CMEs has important consequences for the planet. The implications that will be discussed
in the following are strong atmospheric erosion via stellar wind and CMEs, and the enhanced flux
of galactic cosmic rays onto the planetary atmosphere.

\subsection{Atmospheric escape}\label{sec:consequences:atm}

Here, we explore the region in parameter space where the stellar wind and CMEs can compress the 
magnetosphere down to a certain altitude level. 
In the following we chose 
the critical altitude level to be equal to $R_\text{crit}$=1.15 $R_P$ = 1.15 $R_E$ (i.e.~strong atmospheric loss when
the magnetosphere is compressed to less than $0.15 R_E=1000$ km
above the planetary surface). The numerical study of \citet[][e.g.~Figure 9]{Lammer07AB} shows that 
under certain conditions, the critical altitude level may
be as high as 2 $R_E$ (i.e.~1 $R_E$ above the planetary surface), 
so that this is a conservative assumption.

\begin{figure}
\begin{center}
	\subfigure[Atmospheric erosion by CMEs.]{
 	 \includegraphics[width=0.48\linewidth]{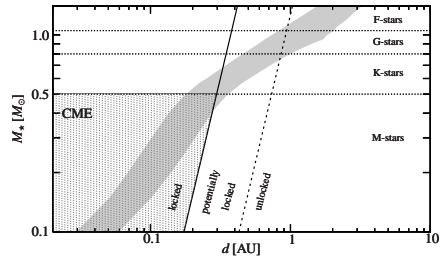}
   \label{fig:CME}}
   	\subfigure[Atmospheric erosion by stellar wind.]{
	 \includegraphics[width=0.48\linewidth]{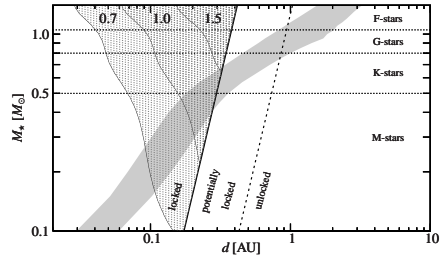}
   	\label{fig:SW}}
\caption{Comparison between the HZ (grey shaded area) and the areas where strong magnetospheric 
compression is possible by CMEs or stellar wind. 
The dotted area denotes the region where the magnetosphere can be compressed down to 1.15
Earth radii or less (i.e., 1000 km above the planetary surface). 
\ref{fig:CME} Magnetospheric compression by strong CMEs. From \citet{Khodachenko07AB}.
\ref{fig:SW} Magnetospheric compression by stellar wind of a young star. 
Three cases are considered: 
Stellar wind of a star of 0.7 Gyr age (all three dotted areas), 
Stellar wind of a star of 1.0 Gyr age (upper two dotted areas),
Stellar wind of a star of 1.5 Gyr age (top dotted area only).
\label{fig:CMEandSW}
}
\end{center}
\end{figure}

Fig.~\ref{fig:CMEandSW} shows
the region in parameter space where
the magnetosphere of an exoplanet with mass, radius, and internal
parameters identical to those of the Earth can be 
strongly compressed  (i.e.~$R_s<R_{\mbox{\scriptsize crit}}$) 
under the action of strong (= dense) CMEs or under the action of a young stellar
wind, respectively. 
In the tidally locked regime, the magnetic moment was taken to be equal to the minimum value, but
tidal locking was assumed only when the timescale for synchronous rotation
is $<100$ Myr (denoted as ``locked''). For timescales between 100 Myr and 10 Gyr (``potentially locked'')
and for timescales above 10 Gyr (``unlocked'') the magnetic moment was estimated using the 
(larger) initial rotation rate of the planet. 

In
Fig.~\ref{fig:CME}, it can be seen that almost all Earth-like exoplanets within habitable zones (HZs, grey shaded 
area) of M-stars will experience this extreme
magnetospheric compression. In such cases, CMEs bring the magnetosphere down to the level of the 
ionopause, and the planetary atmospheres can be
heavily eroded \citep{Khodachenko07AB,Lammer07AB}.

Fig.~\ref{fig:SW} is similar to Fig.~\ref{fig:CME}, but it studies the effect of the
age-dependent stellar wind rather than that of stellar CMEs. 
Three cases are considered: 
The stellar wind of a star of 0.7 Gyr age (all three dotted areas), 
the stellar wind of a star of 1.0 Gyr age (top two dotted areas),
and the stellar wind of a star of 1.5 Gyr age (upper dotted area only).
One can clearly see that for a young star (0.7 Gyr), all planets in a habitable zone between 
0.1 and $\sim$0.3 AU have magnetospheric altitudes lower than the critical altitude level, so that
strong erosion is likely. As the star's age increases, the stellar wind becomes more tenuous, and the
region of strong erosion in the $d$-$M\sstar$-parameter space shrinks. 

An additional effect of stellar age should be mentioned: Not only is the magnetosphere of a close-in 
planet around a young star strongly compressed, but also the planetary atmosphere is more extended 
due to the intense heating by the high stellar X-ray and EUV-flux 
\citep{Lammer03,Griessmeier04, Ribas05}. Thus, the critical level the magnetosphere is allowed to reach,
$R_{\mbox{\scriptsize crit}}(t)$, in reality is age-dependent, and 
it is much more likely for a young planetary magnetosphere
to reach the critical altitude than one would find using present day stellar irradiation.

From Figures \ref{fig:CME} and \ref{fig:SW} it becomes clear that the relatively weak 
intrinsic magnetic fields of tidally locked planets are not suffient to protect Earth-like
exoplanets against significant atmospheric erosion caused by CMEs and the stellar wind. 
The sharp cutoff at
the right edge of the dotted area demonstrates the importance of magnetic shielding: 
Within the dotted area, planets are tidally locked. They are assumed to have a small 
magnetic moment, and magnetic protection is weak. 
To the right of the dotted area however, the planets are assumed to be freely rotating, so that they can
maintain a much higher magnetic moment and are well protected against strong
atmospheric erosion. 

\subsection{Cosmic ray flux}\label{sec:consequences:cr}

\begin{figure}
\begin{center}
 \includegraphics[width=0.48\linewidth]{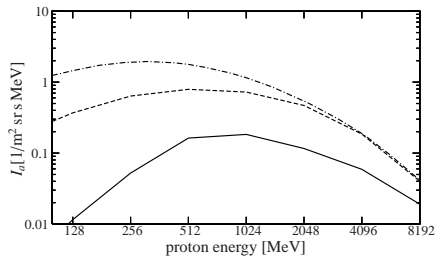} 
 \caption{Dash-dotted line: energy spectrum outside the magnetosphere. 
 	Dashed line: energy spectrum of cosmic ray protons impacting the
	atmosphere (100 km above the surface) of an Earth-like exoplanet at 0.2 AU around a
	0.5M K/M star. Solid line: energy spectrum of cosmic rays impacting the atmosphere
	of the Earth. From \citet{Griessmeier05a}.}
   \label{fig:CR}
\end{center}
\end{figure}

In order to quantify the protection of extrasolar Earth-like planets against galactic cosmic
rays, the motion of galactic cosmic protons through magnetospheres of extrasolar planets has been 
investigated numerically \citep{Griessmeier05a,Griessmeier09}. 
From these calculations, one finds that the cosmic ray flux is strongly
enhanced in the case of a weakly magnetized planet.
Fig.~\ref{fig:CR} shows the reference energy spectrum outside the magnetosphere 
\citep{Seo94} as a dash-dotted line.
The energy spectrum at the top of the Earth's atmosphere is shown as a solid line, and
the cosmic ray energy spectrum for a weakly magnetized Earth-like exoplanet in an orbit of 
0.2 AU around a 0.5 $M_\odot$ K/M star is shown as a dashed line. Due to tidal locking, the 
magnetic moment of that planet is reduced to (at most) 15 \% of the Earth's magnetic moment 
\citep[][section 2.2]{Griessmeier09}, which translates to a strongly enhanced cosmic ray flux
to the planetary atmosphere.

The increased flux of cosmic ray protons of galactic origin to the upper atmospheres of M-star
planets has implications for the flux of (secondary) cosmic ray particles reaching the 
ground, which can be expected to have biological implications.
It also influences the atmospheric composition and thus the remote detection of 
biomarkers as well as the surface UV flux.

\subsubsection{Direct biological impact}\label{sec:consequences:cr:bio}

While a quantitative treatment is not yet available, 
the increased flux of cosmic rays to the top of the planetary atmosphere has 
implications for the flux of secondary radiation to the surface.
When galactic cosmic rays of sufficiently high energy reach the planetary atmosphere, they 
generate showers of secondary cosmic rays.
Of the different components of such showers,
slow neutrons have the strongest influence on biological systems.  
The propagation of high energy cosmic ray particles 
also depends on the composition and density of the planetary atmosphere.
For an Earth-like atmosphere, the minimum energy which a proton must have to initiate a nuclear 
interaction that is detectable at sea level is approximately 450 MeV \citep{Reeves92,Shea00}.
Terrestrial planets with dense atmospheres like Venus (100 bar surface
pressure) would be better shielded by the planetary atmosphere, 
so that no secondary radiation can reach the surface. On the other hand, for planets with thin
atmospheres like Mars (6 mbar surface pressure), 
the surface would probably be totally sterilised, even for a relatively low cosmic ray flux.

Keeping the atmospheric pressure constant at $\sim$1 bar, 
the flux at energies $>450$ MeV is strongly increased for a weakly magnetized planet (see 
Fig.\ref{fig:CR}), and a large increase of  secondary cosmic rays can be expected at the 
planetary surface.

Biological effects, namely an increase in cell fusion indices for different cell-lines, were
found to be significantly correlated with the neutron count rate at the Earth's surface 
\citep{Belisheva05,Griessmeier05a}.
Similar, but much stronger and more diverse effects were observed during large solar particle 
events, where solar cosmic rays dominate over galactic cosmic rays.
The effect of cosmic rays on cells and on DNA are potentially hazardous for life.
On the other hand, changes at the genetic level are a necessary condition for 
biological evolution, so that cosmic rays may also play a role in radiation-induced evolutionary
events.  

As tidally locked Earth-like exoplanets inside the HZ of K/M stars
are only weakly protected against high energetic cosmic
rays, they can be expected to experience a higher surface neutron flux and stronger biological
effects than Earth-like planets with a strong magnetic field.
For this reason, it may be more difficult for life to develop on the surface of 
planets around such stars.

\subsubsection{Atmospheric modification}\label{sec:consequences:cr:atm}

The influence of a high flux of galactic cosmic rays (GCRs) and of stellar energetic particles
on the atmospheres of planets in the HZ
around M-stars is studied by \citet{Grenfell07AB} and \citet{Grenfell09}. 
When cosmic ray particles travel through an Earth-like atmosphere, they have sufficient energy to
break 
the strong $\mathrm{N_2}$ molecule, and the chemical products react with oxygen to form nitrogen 
oxides ($\mathrm{NO_x}$).  If UV levels are intense, e.g. corresponding to the upper mesosphere
and above on the Earth, then the $\mathrm{NO_x}$ cannot survive. $\mathrm{NO_x}$ affects
biomarkers differently depending on altitude. 
In a stratospheric environment, $\mathrm{NO_x}$ depletes ozone by catalytic cycles \citep{Crutzen70}. Once
ozone is affected, the other biomarkers can change too, because ozone affects the temperature
profile (hence e.g. mixing processes, chemical sources and sinks) and UV levels (hence
photolysis rates) for other chemical species. 

This is important, because 
atmospheres of extrasolar planets will
be studied from the point of view of possible biological
activity in the near future. ``Atmospheric biomarkers'' are compounds present
in an atmosphere which imply the presence of life and
which cannot be explained by inorganic chemistry alone.
The simultaneous presence of significant amounts of
atmospheric reducing gases (e.g. methane) and oxidising gases
(e.g. oxygen) as on the Earth is also an indicator
of biological activity \citep{Lovelock65,Sagan93}.
Good biomarkers include oxygen (produced by photosynthesis),
ozone (mainly produced from oxygen) and nitrous oxide
(produced almost exclusively from bacteria).
For the study of biomarkers in planetary atmospheres, it is important
to know all inorganic effects which can reduce the abundances of these
molecules.
Modelling studies 
are required to rule out false conclusions in cases where inorganic
chemistry, e.g.~reactions triggered by the influx of cosmic ray particles, 
can mimic (or mask) the presence of life. 

\begin{figure}
\begin{center}
 \includegraphics[width=0.3\linewidth,angle=270]{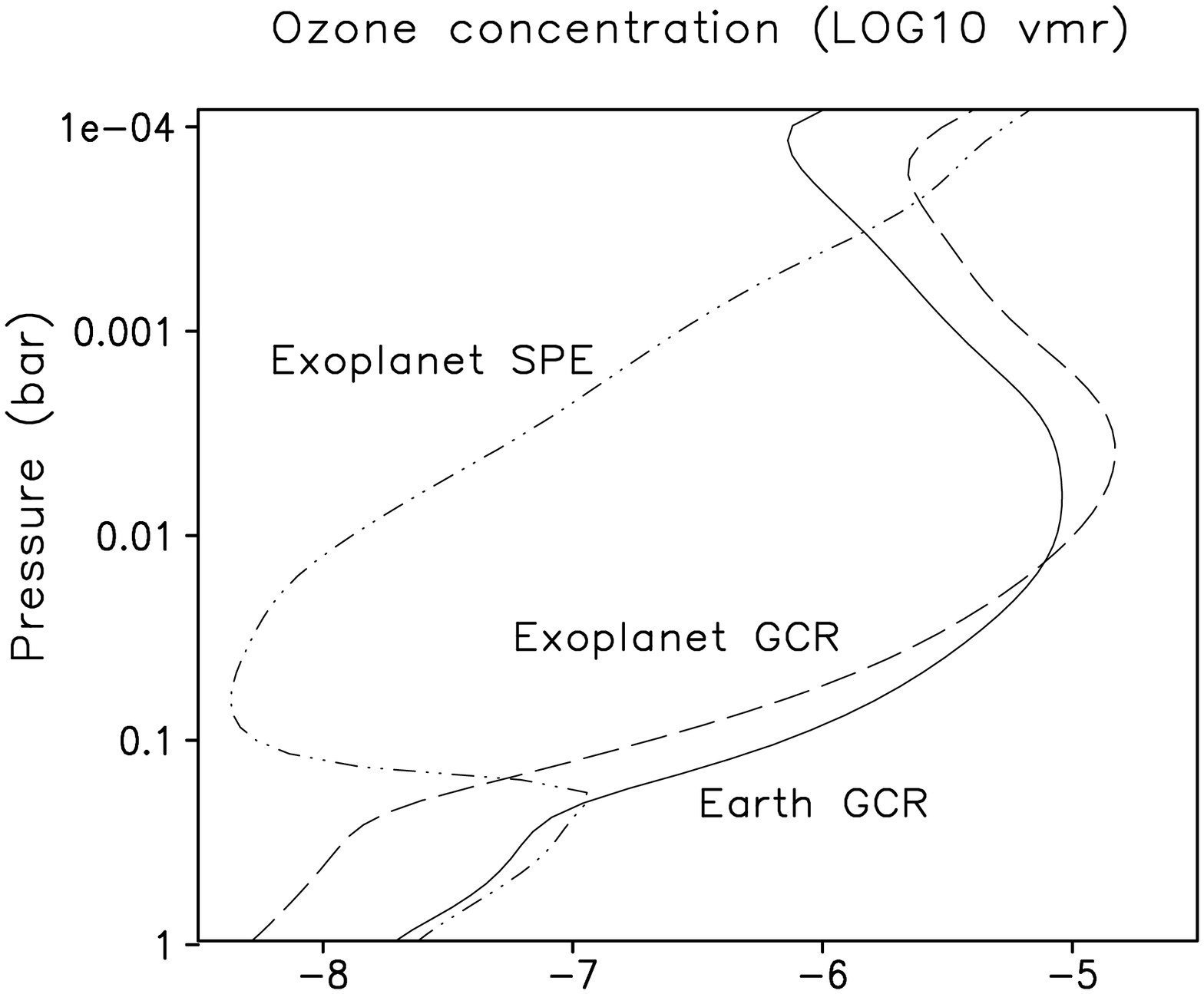} 
 \caption{Atmospheric biomarker profile concentrations (log volume mixing ratio): ozone for
 	 three different cases (see text).  From \citet[][]{Grenfell09}.}
   \label{fig:O3}
\end{center}
\end{figure}

Using the cosmic ray spectrum of Fig.~\ref{fig:CR}, \citet{Grenfell07AB} found 
that for Earth-like planets in close orbits around M-stars,
GCRs
may only change the abundances of biomarker molecules (especially water and ozone) by 
a small factor.
This can also be seen by comparing the cases ``Earth GCR'' and ``Exoplanet GCR'' in Fig.~\ref{fig:O3}.
The case is different for solar energetic particles \citep{Grenfell09},
which can modify the abundance of ozone in the planetary atmosphere by several orders of magnitude.
This effect can be seen in Fig.~\ref{fig:O3}, where ``Exoplanet SPE'' shows the effect of solar 
energetic particles \citep[][]{Grenfell09}.
Such changes in atmospheric biomarker concentrations have to be taken into account when 
searching for biosignatures in the spectra of Earth-like exoplanets 
by future space missions like DARWIN \citep{Fridlund04} or SEE-COAST \citep{Schneider06}. 

\subsubsection{Modification of UV shielding}\label{sec:consequences:cr:uv}

Both direct UV radiation
\citep{Buccino07,Scalo07} and indirect UV radiation generated by energetic particles \citep{Smith04} 
can have important consequences for biogenic processes on M-star planets.
As was shown in Section \ref{sec:consequences:cr:atm}, the planetary ozone layer of a weakly 
magnetized planet can be virtually destroyed during a solar energetic particle event.
The corresponding enhancement of the UV flux has to be taken into account when biological 
effects of UV radiation are considered.
The influence of UV radiation of different wavelenth on life is also discussed by 
\citet{Cuntz10}.

\section{Conclusions}\label{sec:conclusions}

Because of tidal locking, extrasolar planets with orbits in the habitable zone of K/M are likely to
have much smaller magnetic moments than more distant planets. Combined with the strong ram pressure
of dense and fast stellar winds and the high CME activity expected around young stars, this leads to
the conclusion that such planets will have small magnetospheres which offer only limited magnetic
protection.

The weak magnetic protection may lead to strong atmospheric erosion during the early stages of
stellar evolution, which could be strong enough to erode the atmosphere or the planets' water 
inventory via non-thermal atmospheric loss processes.
Another consequence of the weak magnetic protection is a high flux of galactic cosmic rays and
stellar energetic particles to the planetary atmosphere. 
This has implications for potential habitability, but also for the atmospheric chemistry and
composition (e.g.~destruction of atmospheric ozone). 
These effects have also to be considered for missions studying biosignatures in the observed
spectra of Earth-like exoplanets.

Overall, the magnetic protection of exoplanets in the habitable zone of K/M stars can be weak, so
that these planets may only be weakly shielded against stellar activity.

{\footnotesize 
\section{Acknowledgements}

This study was supported by the International Space Science
Institute (ISSI) through the ISSI Team
``Evolution of Exoplanet Atmospheres and their Characterisation'' 
and by the Helmholtz association through the research alliance
``Planetary Evolution and Life''. 
MK acknowledges the Austrian Fond zur F\"{o}rderung der wissenschaftlichen Forschung (project 
P21197-N16). 


\end{document}